\documentstyle[aps,preprint,eqsecnum,floats,epsf]{revtex}
\tighten
\draft
\begin{document}

\preprint{}
\title{Barrier Penetration for Supersymmetric Shape-Invariant
Potentials}
\author{A.~N.~F. Aleixo\thanks{Electronic address:
        {\tt aleixo@nucth.physics.wisc.edu}}}
\address{Department of Physics, University of Wisconsin\\
         Madison, Wisconsin 53706 USA,\\
         Instituto de F\'{\i}sica, Universidade Federal 
         do Rio de Janeiro, RJ - 
         Brazil\thanks{{\tt Permanent address.}}}
\author{A.~B. Balantekin\thanks{Electronic address:
        {\tt baha@nucth.physics.wisc.edu}},}
\address{Department of Physics, University of Wisconsin\\
         Madison, Wisconsin 53706 USA}
\author{M.~A. C\^andido Ribeiro\thanks{Electronic address:
         {\tt macr@df.ibilce.unesp.br}},}
\address{Departamento de F\'{\i}sica - Instituto de Bioci\^encias,
         Letras e Ci\^encias Exatas\\
         UNESP, S\~ao Jos\'e do Rio Preto, SP - Brazil}
\date{\today}
\maketitle

\begin{abstract}
  Exact reflection and transmission coefficients for supersymmetric
  shape-invariant potentials barriers are calculated by an analytical
  continuation of the asymptotic wave functions obtained via the
  introduction of new generalized ladder operators. The general form
  of the wave function is obtained by the use of the {\bf
    F}$(-\infty,+\infty)$-matrix formalism of Fr\"oman and Fr\"oman
  which is related to the evolution of asymptotic wave function
  coefficients.
\end{abstract}

\pacs{}

\newpage
\section{Introduction}

Quantum tunneling through a potential barrier governs many interesting
phenomena in physics ranging from fusion reactions in stars
\cite{reference1} to the study of transitions from metastable states
\cite{reference2}. There are very few exactly solvable examples of
barrier penetration. Supersymmetric quantum mechanics has been shown
to be a useful technique to explore exactly solvable problems in
quantum mechanics \cite{ref1}.  An integrability condition called
shape-invariance was introduced by Gendenshtein \cite{ref2} and was
cast into an algebraic form by Balantekin \cite{ref3}. Reflection and
transmission coefficients for a large class of shape-invariant
potentials was given by Cooper et al \cite{coopgin}. A general
operator method for calculating scattering amplitudes for
supersymmetric shape-invariant potentials was introduced by Khare and
Sukhatme \cite{kharesuk}. Even though an approximate method in the
context of supersymmetric semiclassical approximation \cite{susywkb1}
to calculate tunneling through one-dimensional potential barriers was
presented in \cite{susywkbtun} exact tunneling probabilities for
shape-invariant barriers was not explicitly derived. We cover this
latter subject in this article.

 Introducing the superpotential function
\begin{equation}
W(x) \equiv
-\frac{\hbar}{\sqrt{2m}}\left[ \frac{\Psi^\prime_0(x)}
{\Psi_0(x)}\right] \,,
\label{sup}
\end{equation}
where $\Psi_0(x)$ is the ground-state wave function of the Hamiltonian
$\hat H$, and defining the operators 
\begin{equation}
\hat A \equiv W(x) + \frac{i}{\sqrt{2m}}\hat p\,,
\label{opa}
\end{equation}
\begin{equation}
\hat A^\dagger \equiv W(x) - \frac{i}{\sqrt{2m}}\hat p\,,
\label{opad}
\end{equation}
we can show that

\begin{equation}
\hat H - E_0 = \hat A^\dagger \hat A\,.
\label{eqhe}
\end{equation}
Since the ground-state wave function satisfies the condition 

\begin{equation}
\hat A \Psi_0(x) = 0
\label{eqgs}
\end{equation}
the supersymmetric partner potentials 

\begin{equation}
{\hat H}_1 = {\hat A}^\dagger{\hat A}\,\qquad {\hat H}_2 = {\hat
A}{\hat A}^\dagger
\label{eqh12}
\end{equation}
have the same energy spectra except the ground state of ${\hat
H}_1$ which has no corresponding state in the spectra of ${\hat
H}_2$. The corresponding potentials are given by 

\begin{equation}
V_1(x) = \left[W(x)\right]^2-{\hbar\over\sqrt{2m}}{dW\over dx}
\label{eqv1}
\end{equation}

\begin{equation}
V_2(x) = \left[W(x)\right]^2+{\hbar\over\sqrt{2m}}{dW\over dx}
\label{eqv2}
\end{equation}
The shape-invariance condition\cite{ref2}

\begin{equation}
V_2(x,a_1) = V_1(x,a_2) + R(a_1)
\label{eqvv12}
\end{equation}
can also be written as\cite{ref3} 
\begin{equation}
\hat A(a_1) \hat A^\dagger(a_1) =\hat A^\dagger (a_2) 
\hat A(a_2) + R(a_1) \,,
\label{eqsi}
\end{equation}
where $a_{1,2}$ are a set of parameters that specify space-independent
properties of the potentials (such as strength, range, and
diffuseness). The parameter $a_2$ is a function of $a_1$ and the
remainder $R(a_1)$ is independent of $\hat x$ and $\hat p$.  Not all
exactly solvable potentials are shape-invariant \cite{ref4}. In the
cases studied so far the parameters $a_1$ and $a_2$ are either related
by a translation \cite{ref4,ref5} or a scaling \cite{ref6}.
Introducing the similarity transformation that replaces $a_1$ with
$a_2$ in a given operator
\begin{equation}
\hat T(a_1)\, \hat O(a_1)\, \hat T^\dagger(a_1) = \hat O(a_2)
\label{eqsio}
\end{equation}
and the operators
\begin{equation}
\hat B_+ =  \hat A^\dagger(a_1)\hat T(a_1)
\label{eqba}
\end{equation}
\begin{equation}
\hat B_- =\hat B_+^\dagger =  \hat T^\dagger(a_1)\hat A(a_1)\,,
\label{eqbe}
\end{equation}
the Hamiltonian takes the form
\begin{equation}
\hat H - E_0 =\hat B_+\hat B_-\,.
\label{eqhb}
\end{equation}
Using Eq.~(\ref{eqsi}) one can easily prove the commutation relation

\begin{equation}
[\hat B_-,\hat B_+] =  \hat T^\dagger(a_1)R(a_1)\hat T(a_1) 
\equiv R(a_0)\,,
\label{eqcb1}
\end{equation}
where we used the identity

\begin{equation}
R(a_n) = {\hat T}(a_1)\,R(a_{n-1})\,{\hat T}^\dagger (a_1)\,,
\label{eqran}
\end{equation}
valid for any $n$. Equation (\ref{eqcb1}) suggests that ${\hat B}_+$
and ${\hat B}_-$ are the appropriate creation and annihilation
operators provided that their non-commutativity with $R(a_1)$ is taken
into account.  In this paper we extend the use of ${\hat B}_+$ and
${\hat B}_-$ operators for calculating the asymptotic behaviors of the
wave functions related with incidence of a particle on a
supersymmetric shape-invariant potential barrier and obtain the exact
transmission and reflection coefficients.

\section{Exact Wave functions}

The wave functions for all currently known supersymmetric
shape-invariant potential barriers can be calculated analytically
using supersymmetric operator techniques\cite{ref9,ref10}.  The final
result can be expressed by single operators $\hat B_+$ and $\hat
B_-^{-1}$\cite{ref11} or by couples of these operators. In the first
case we can use the two additional commutation relations

\begin{equation}
[\hat B_+\hat B_-,\hat B_+^n] =
\sum_{k=1}^n R(a_k)\,\hat B_+^n\,,
\label{eqb+n}
\end{equation}
and

\begin{equation}
[\hat B_+\hat B_-,\hat B_-^{-n}] =
\sum_{k=1}^n R(a_k)\,\hat B_-^{-n}\,,
\label{eqb-n}
\end{equation}
obtained by induction using the relations 

\begin{equation}
R(a_n)\hat B_+ = \hat B_+R(a_{n-1})\,,
\label{eqran+}
\end{equation}

\begin{equation}
R(a_n)\hat B_- = \hat B_+R(a_{n+1})\,,
\label{eqran-}
\end{equation}
that readily follow from Eqs. (\ref{eqsio}), (\ref{eqba}) and
(\ref{eqbe}).  Considering that the Schr\"odinger equation can be
write as

\begin{equation}
\hat B_+\hat B_-\Psi (x) = \Lambda \Psi(x)\,,
\label{eqsch}
\end{equation}
then the Eqs. (\ref{eqb+n}) and (\ref{eqb-n}) imply that $\hat B_+$
and $\hat B_-$ can be used as ladder operators to solve the
Eq.~(\ref{eqsch})\cite{ref3,refbma}. To this end we introduce
$\Psi_-^{(0)}(x)$ as the solution of the equation

\begin{equation}
\hat A_-(a_1)\,\Psi_-^{(0)}(x) = 0 = \hat B_-(a_1)\,\Psi_-^{(0)}(x)\,,
\label{eqaps0}
\end{equation}
which implies

\begin{equation}
\Psi_-^{(0)}(x, a_1) \sim 
\exp{\left( -{\sqrt{2m}\over\hbar}\int^x d\xi \,
W(\xi,a_1)\right) }\,.
\label{eqps0-}
\end{equation}
If the function 

\begin{equation}
f(n) = \sum_{k=1}^n R(a_k)
\label{eqsr}
\end{equation}
can be analytically continued so that the condition 

\begin{equation}
f(\mu) = \Lambda
\label{eqfl}
\end{equation}
is satisfied for a particular (in general complex) value of $\mu$,
then the Eq.~(\ref{eqb+n}) implies that one possible form for the
solution of Eq.~(\ref{eqsch}) is \ $\hat B_+^\mu\Psi_-^{(0)}(x,a_1)$\ 
.  Similarly if $\Psi_+^{(0)}(x)$ satisfies the equation

\begin{equation}
\hat B_+(a_1)\,\Psi_+^{(0)}(x) = 0 \,,
\label{eqbps0}
\end{equation}
which implies that

\begin{equation}
\hat T(a_1)\Psi_+^{(0)}(x) \sim \exp{\left( 
{\sqrt{2m}\over\hbar}\int^x d\xi \,
W(\xi,a_1)\right) }\,,
\label{eqtps0+}
\end{equation}
or

\begin{equation}
\Psi_+^{(0)}(x,a_0) \sim \exp{\left( {\sqrt{2m}\over\hbar}\int^x 
d\xi \, W(\xi,a_0)\right) }\,,
\label{eqps0+}
\end{equation}
then the Eq.~(\ref{eqb-n}) implies that other possible form for the
solution of Eq.~(\ref{eqsch}) is \ $\hat
B_-^{-\mu-1}\Psi_+^{(0)}(x,a_0)$\ .  At this point we conclude that
the components of the wave functions, written in terms of the singles
operators $\hat B_+$ and $\hat B_-^{-1}$, for supersymmetric
shape-invariant potential barriers can be written down as

\begin{mathletters}
\label{eqpsso}
\begin{eqnarray}
\Psi_-(x) &=& \beta  \hat B_+^\mu\Psi_-^{(0)}(x,a_1) \\
\Psi_+(x) &=& \gamma \hat B_-^{-\mu-1}\Psi_+^{(0)}(x,a_0) \,,
\end{eqnarray}
\end{mathletters}
where $\mu$ is obtained by the relation
\begin{equation}
\Lambda =  \sum_{k=1}^\mu R(a_k)
\label{eqlda}
\end{equation}
where $\beta$ and $\gamma$ are constants and 

\begin{equation} 
\Psi_\pm^{(0)}(x,a_\mu) = \exp{\left( \pm 
{\sqrt{2m}\over\hbar}\int^x d\xi \,
W(\xi,a_\mu )\right) }\,.
\label{eqpm0}
\end{equation}
With each value of $\mu$ we obtain several possible expressions for
the components $\Psi_\pm (x)$ and the general expression for the wave
function can be obtained with these components or a combination of
them.
 
It is also possible to express the components of the wave functions
using couples of the operators $\hat B_+$ and $\hat B_-^{-1}$.  In
this case we can use the equations (\ref{eqb+n}), (\ref{eqb-n}) and
the relations (\ref{eqran+}), (\ref{eqran-}) to show by induction that

\begin{equation}
[\hat B_+\hat B_-,(\hat B_+\hat B_-^{-1})^n] =
\sum_{k=1}^{2n} R(a_k)\,(\hat B_+\hat B_-^{-1})^n\,,
\label{eqb+b-n}
\end{equation}
and
\begin{equation}
[\hat B_+\hat B_-,(\hat B_-^{-1}\hat B_+)^n] =
\sum_{k=1}^{2n} R(a_k)\,(\hat B_-^{-1}\hat B_+)^n\,.
\label{eqb-b+n}
\end{equation}

Using these last two equations and the same conditions (\ref{eqaps0})
and (\ref{eqbps0}) we can show that the components of the wave
functions, written in terms of couples of the operators $\hat B_+$ and
$\hat B_-^{-1}$, for supersymmetric shape-invariant potential barriers
can be written down as

\begin{mathletters}
\label{eqps+-}
\begin{eqnarray}
\Psi_-(x) &=& \beta \, 
(\hat B_+\hat B_-^{-1})^\nu\,\Psi_-^{(0)}(x,a_1) \\
\Psi_+(x) &=& \gamma 
\,(\hat B_-^{-1}\hat B_+)^\nu\hat B_-^{-1}\,\Psi_+^{(0)}(x,a_0)
\end{eqnarray}
and 
\end{mathletters}
where $\nu$ is obtained by the relation
\begin{equation}
\Lambda =  \sum_{k=1}^{2\nu} R(a_k)\,.
\label{eqllda}
\end{equation}
Note that in a given problem either Eqs. (\ref{eqpsso}) or
Eqs. (\ref{eqps+-}) could be used, but not both. 


\section{Asymptotic Wave functions}

The formal expressions for the components of the wave functions can be
express in explicit forms if we evaluate them asymptotically. In the
case of single operators expressions we first note that using Eqs.
(\ref{eqba}) and (\ref{eqbe}) the results (\ref{eqpsso}) can be
written as

\begin{mathletters}
\label{eqpssa}
\begin{eqnarray}
\Psi_-(x) &=& \beta  \hat A_+(a_1)\hat A_+(a_2)\cdots  \hat A_+(a_\mu
       )\Psi_-^{(0)}(x,a_{\mu +1}) \\
       \Psi_+(x) &=& \gamma \hat A_-^{-1}(a_1)\hat A_-^{-1}(a_2)
       \cdots \hat A_-^{-1}(a_{\mu+1}
        )\Psi_+^{(0)}(x,a_{\mu+1} )
\end{eqnarray}
\end{mathletters}
In this point we need to consider the two basic asymptotic behavior
for the superpotential: i) $W(x\rightarrow \pm\infty ,a_\mu )$ is
constant (i.e., the potential barrier goes to a constant); and ii)
$W(x\rightarrow\pm\infty ,a_\mu )\longrightarrow\pm\infty\,$ (i.e.,
the potential barrier goes to $-\infty$).  In the former limit the
commutator

\begin{equation}
\left[ {\partial\over\partial x},W(x,a_\mu)\right] = 
W^\prime (x,a_\mu )
\label{eqcw}
\end{equation}
vanishes. In the ladder case this commutator can be ignored as 

\begin{eqnarray}
W(x,a_n)\,W(x,a_k) + W^\prime(x,a_n) &=& W(x,a_n)\,W(x,a_k)\left( 1
+ {W^\prime(x,a_n)\over W(x,a_n)\,W(x,a_k)}\right) 
\nonumber \\
&\rightarrow & W(x,a_n)\,W(x,a_k)\,,
\label{eqwe}
\end{eqnarray}
provided that $W^\prime (x,a_n)/W(x,a_n)$ remains finite, which is the
case for all realistic superpotentials. Hence in both limits we can
write Eqs.~(\ref{eqpssa}) as

\begin{mathletters}
\label{eqpssd}
\begin{eqnarray}
\Psi_-(x) &=& \beta
(W_1+W_{\mu +1})(W_2+W_{\mu +1})\cdots (W_\mu +W_{\mu +1})\,
\Psi_-^{(0)}(x,a_{\mu +1})\\
      \Psi_+(x) &=& \gamma
(W_1+W_{\mu+1} )^{-1}(W_2+W_{\mu+1} )^{-1}\cdots (W_{\mu+1}
+W_{\mu+1} )^{-1}
\,\Psi_+^{(0)}(x,a_{\mu+1} )\,.
\end{eqnarray}
\end{mathletters}
In these equations the quantity $W_m$ is the short-hand notation for
$W(x,a_m)$. 

If we assume that the superpotential satisfies the condition

\begin{equation}
W(x,a_n) = W(x,a_1) + (n-1)\,\zeta (x)\,,
\label{eqwz}
\end{equation}
then these asymptotic equations can be in a form suitable for analytic
continuation, that is

\begin{mathletters}
\label{eqpssf}
\begin{eqnarray}
\Psi_-(x) &=& \beta\,\zeta ^\mu\, {\Gamma (2z+2\mu)\over \Gamma
(2z+\mu )}\,\Psi_-^{(0)}(x,a_{\mu +1}) \\ 
      \Psi_+(x) &=& \gamma\,\zeta ^{-\mu -1}\, {\Gamma (2z+\mu)\over \Gamma
(2z+2\mu +1 )}\,\Psi_+^{(0)}(x,a_\mu )\,,
\end{eqnarray}
\end{mathletters}
where $z = W(x,a_1)/\zeta (x)$.  The condition given by
Eq.~(\ref{eqwz}) is satisfied for a number of superpotentials and in
the last section we give some examples. If this condition is not
satisfied, the analytic continuation may still be done, but will be
more complicated.

When the asymptotic behavior of the superpotential is $W(x,a_n)
\rightarrow \pm\infty$ we can use the identity

\begin{equation}
\lim_{y\to\pm\infty}\,{1\over y^\mu}{\Gamma(y+2\mu
)\over\Gamma(y+\mu)} = 1
\label{eqlimg}
\end{equation}
to express Eqs.~(\ref{eqpssf}) in the simple form

\begin{mathletters}
\label{eqpssl}
\begin{eqnarray}
\Psi_-(x) &=& \beta\,(2W_1) ^\mu\,\Psi_-^{(0)}(x,a_{\mu +1})\\
      \Psi_+(x) &=& \gamma\,(2W_1) 
^{-\mu -1}\,\Psi_+^{(0)}(x,a_{\mu +1})\,.
\end{eqnarray}
\end{mathletters}

We can repeat the same procedure used above in the case of couple of
operators. Again, using Eqs.~(\ref{eqba}) and (\ref{eqbe}) the results
(\ref{eqps+-}) can be written as

\begin{mathletters}
\label{eqpsa}
\begin{eqnarray}
\Psi_-(x) &=& \beta\,\prod_{k=1}^\nu\, \hat A_+(a_{2k-1})\,
\hat A_-^{-1}(a_{2k})\,\Psi_-^{(0)}(x,a_{2\nu +1})\\
  \Psi_+(x) &=& \gamma\,\prod_{k=1}^\nu\, \hat A_-^{-1}(a_{2k-1})\,
\hat A_+(a_{2k})\,\hat A_-^{-1}(a_{2\nu +1})\,
\Psi_+^{(0)}(x,a_{2\nu +1})\,.
\end{eqnarray}
\end{mathletters}
Considering the superpotential asymptotic simplifications given by
Eqs.~(\ref{eqcw}), (\ref{eqwe}) and the analytic continuation
condition (\ref{eqwz}) we can write the result for the components of
the asymptotic wave functions in this case as

\begin{mathletters}
\label{eqpsb}
\begin{eqnarray}
\Psi_-(x) &=& \beta\,{\Gamma (1-z-\nu)\over \Gamma (1-z-2\nu )\,
\Gamma (\nu + {1\over 2})}\,\Psi_-^{(0)}(x,a_{2\nu +1})\\
    \Psi_+(x) &=& \gamma\,{\Gamma (-z-2\nu )\,\Gamma (\nu +{1\over2})
    \over \Gamma (1-z-\nu )}\,\Psi_+^{(0)}(x,a_{2\nu +1})\,.
\end{eqnarray}
\end{mathletters}


\section{General Asymptotic Wave functions and the Transmission and
Reflection Coefficients}

Using the formalism developed in the Ref.~\cite{ref16} we can write
two possible asymptotic solutions for the one-dimensional
time-independent Schr\"odinger equation in the form

\begin{mathletters}
\label{eqfa1}
\begin{eqnarray}
\Psi_1(x\rightarrow\pm\infty) &=& A_{11}(\pm\infty)\,f_1(x) + 
                                       A_{21}(\pm\infty)\,f_2(x)\\
       \Psi_2(x\rightarrow\pm\infty) &=& A_{12}(\pm\infty)\,f_1(x) + 
                                       A_{22}(\pm\infty)\,f_2(x)\,,
\end{eqnarray}
\end{mathletters}
where 

\begin{equation}
f_1(x) = {\exp({+i\chi (x)})\over\sqrt{q(x)}}\qquad {\rm and}\qquad
       f_2(x) = {\exp({-i\chi (x)})\over\sqrt{q(x)}}\,,
\label{eqfa2}
\end{equation}
with
\begin{equation}
\chi (x) = \int^x q(\xi )\,d\xi\qquad{\rm and}\qquad
      q(x) = {\sqrt{2m}\over \hbar}\,W(x,a_{\mu+1})\,.
\label{eqfa3}
\end{equation}
If we define the vectors 

\begin{equation}
\mid \Psi (x)\rangle = \left[ \matrix{ 
\mid \Psi_1 (x) \rangle \cr \mid \Psi_2 (x) \rangle \cr}\right]
\label{eqfa4}
\end{equation}
and

\begin{equation}
\langle f(x)\mid\,  = \left[ \matrix{ \langle f_1 (x) \mid \quad
    \langle f_2 (x) \mid \cr}\right]
\label{eqfa5}
\end{equation}
then we can write 
\begin{equation}
\langle \Psi (x\rightarrow\pm\infty )\mid = \langle f(x)\mid {\bf
A}(\pm\infty)\,,
\label{eqfa6}
\end{equation}
where the asymptotic coefficients matrix is given by

\begin{equation}
{\bf A}(\pm\infty ) = \left[ \matrix{A_{11}(\pm\infty)
&A_{12}(\pm\infty)\cr A_{21}(\pm\infty)&A_{22}(\pm\infty)
\cr}\right]
\label{eqfa7}
\end{equation}
Considering that the space evolution of the coefficients matrix 
{\bf A}, given by an iteration process, can be written as

\begin{equation}
{\bf A}(x) = {\bf F}(x,x_0)\,{\bf A}(x_0)\,,
\label{eqfa8}
\end{equation}
then if we know the asymptotic coefficients in $-\infty$ and $+\infty$
we can obtain the evolution matrix

\begin{equation}
{\bf F}(-\infty ,+\infty) = {\bf A}(-\infty )\,{\bf
A}^{-1}(+\infty)\,.
\label{eqfa9}
\end{equation}
Using the Eq.~(\ref{eqfa9}) we can show two basics properties of the
{\bf F} matrix:

\begin{mathletters}
\label{eqpfpr}
\begin{eqnarray}
{\rm det}\, {\bf F}(-\infty,+\infty) &=& 1 \\  
{\bf F}(-\infty,+\infty) &=& {\bf F}^{-1}(+\infty,-\infty)\,.
\end{eqnarray}
\end{mathletters}
The knowledge of {\bf F}$(-\infty,+\infty)$ permits the determination
of the exact transmission and reflection barrier coefficients.  If we
consider a wave incidence from $-\infty$ to $+\infty$ then we can
write the asymptotic wave function in the form

\begin{mathletters}
\label{eqfa11}
\begin{eqnarray}
\Psi(x\rightarrow-\infty) &=& f_1(x) + C_R\,f_2(x)\\
       \Psi(x\rightarrow+\infty) &=& C_T\,f_2(x) \,,
\end{eqnarray}
\end{mathletters}
or

\begin{mathletters}
\label{eqfa12}
\begin{eqnarray}
\Psi(x\rightarrow-\infty) &=& \langle f(x)\mid a(-\infty)\rangle \\
\Psi(x\rightarrow+\infty) &=& \langle f(x)\mid a(+\infty)\rangle  \,,
\end{eqnarray}
\end{mathletters}
where

\begin{equation}
\mid a(-\infty) \rangle = \left[ \matrix{ 1\cr
C_R\cr}\right]\qquad{\rm and}\qquad
\mid a(+\infty) \rangle = \left[ \matrix{ 0\cr
C_T\cr}\right]\,.
\label{eqfa13}
\end{equation}
Considering that

\begin{eqnarray}
\mid a(-\infty) \rangle = {\bf F}(-\infty,+\infty)\,\mid
a(+\infty)\rangle
\label{eqfa14}
\end{eqnarray}
we can conclude that

\begin{mathletters}
\label{eqfa15}
\begin{eqnarray}
C_T &=& {1\over F_{12}(-\infty,+\infty)} \\
 C_R &=& {F_{22}(-\infty,+\infty)\over F_{12}(-\infty,+\infty)}\,,
\end{eqnarray}
\end{mathletters}
and the transmission and reflection coefficients are given by $T =
\mid C_T\mid^2$ and $R = \mid C_R\mid ^2$.  At this point if we
consider the conservation of probability, the time reversal
invariance, and the invariance under space reflection it is possible
to show the additional {\bf F}-matrix properties:

\begin{mathletters}
\label{eqfa16}
\begin{eqnarray}
F_{11}(-\infty,+\infty) &=& -F_{22}(-\infty,+\infty) \\
       F_{21}(-\infty,+\infty) &=& -F_{12}(-\infty,+\infty) \\
       \mid F_{12}(-\infty,+\infty)\mid\,  &=&\, \mid
       F_{12}(+\infty,-\infty)\mid\ge 1 \\
       \mid F_{22}(-\infty,+\infty)\mid\,  &=& \,\mid
       F_{22}(+\infty,-\infty)\mid\le\mid
       F_{12}(-\infty,+\infty)\mid\,.       
\end{eqnarray}
\end{mathletters}


\section{Applications}

\subsection{Parabolic Barrier}

For a parabolic potential barrier\cite{refbellp}

\begin{equation}
V_1(x) = V_0 - {1\over 2}m\Omega^2x^2\,,
\label{eqpb1}
\end{equation}
the corresponding superpotential, obtained by using the
Eq.~(\ref{eqv1}), is given by

\begin{equation}
W(x,a_1) = a_1x
\label{eqwp1}
\end{equation}
where $a_1 = \pm i\sqrt{m\over 2}\Omega$. The shape invariance
condition (\ref{eqvv12}) imply that

\begin{equation}
R(a_n) = \pm 2i\varepsilon_0
\label{eqran1}
\end{equation}
where $\varepsilon_0 = {1\over 2}\hbar\Omega$. Using the 
Eq.~(\ref{eqllda}) we can conclude that

\begin{equation}
\sum_{k=1}^\mu R(a_k) = \pm i2\mu\varepsilon_0 = \Lambda = E-V_0\mp
i\varepsilon_0
\label{eqrl}
\end{equation}
or

\begin{equation}
\mu = -{1\over 2}\pm i{\lambda\over 2}
\label{eqmup}
\end{equation}
where $\lambda = (V_0 - E)/\varepsilon_0$.
The asymptotic form for the components of the wave function can be
obtained using these results in the Eqs.~(\ref{eqpssl})

\begin{mathletters}
\label{eqpas}
\begin{eqnarray}
\Psi_-(x) &=& \beta \,(\pm i)^{-{1\over 2}\pm i{\lambda\over 2}}\,
\varrho^{\pm i{\lambda\over 2}} \left[{\exp{(\mp i\varrho^2)}\over
\sqrt{\varrho}}\right] \\
\Psi_+(x) &=& \gamma\,(\pm i)^{-{1\over 2}\mp i{\lambda\over 2}}\,
\varrho^{\mp i{\lambda\over 2}} \left[{\exp{(\pm i\varrho^2)}\over
\sqrt{\varrho}}\right]\,
\end{eqnarray}
\end{mathletters}
or

\begin{equation}
\Psi_-(x) = \beta\left\{\matrix{ 
e^{\mp i\left( k+{1\over
4}\right)\pi}\,e^{-\left( k+{1\over 4}\right)\pi\lambda}\cr 
\cr
e^{\pm
i\left( k +{3\over 4}\right)\pi}\,e^{\left( k+{3\over
4}\right)\pi\lambda}\cr}\right\}\,
\varrho^{\pm i{\lambda\over 2}}\,\left[{\exp( {\mp
i\varrho^2})\over \sqrt{\varrho}}\right]\,,
\label{eqpas1}
\end{equation}
and

\begin{equation}
\Psi_+(x) = \gamma\left\{\matrix{ 
e^{\mp i\left( k+{1\over
4}\right)\pi}\,e^{\left( k+{1\over 4}\right)\pi\lambda}\cr 
\cr
e^{\pm
i\left( k +{3\over 4}\right)\pi}\,e^{-\left( k+{3\over
4}\right)\pi\lambda}\cr}\right\}\,
\varrho^{\mp i{\lambda\over 2}}\,\left[ {\exp({\pm
i\varrho^2})\over \sqrt{\varrho}}\right]\,,
\label{eqpas2}
\end{equation}
where \ $\varrho = \sqrt{m\Omega/2\hbar}\,x$\ and \ $k =
0,1,2,\ldots$\ .  Using these results it is possible to write two
asymptotic solutions of the Schr\"odinger equation when $x\rightarrow
+\infty$ in the form

\begin{mathletters}
\label{eqppai}
\begin{eqnarray}
\Psi_1(x\rightarrow +\infty) &=& e^{i{\pi\over
4}}e^{-{\pi\lambda\over 4}}\mid \varrho\mid^{i{\lambda\over
2}}\,\left[ {\exp({-i\varrho^2})\over\sqrt{\varrho}}\right] 
+ e^{-i{\pi\over
4}}e^{-{\pi\lambda\over 4}}\mid \varrho\mid^{-i{\lambda\over
2}}\,\left[ {\exp({+i\varrho^2})\over\sqrt{\varrho}}\right] \\
\Psi_2(x\rightarrow +\infty) &=&  e^{-i{\pi\over
4}}e^{-{3\pi\lambda\over 4}}\mid \varrho\mid^{i{\lambda\over
2}}\,\left[ {\exp({-i\varrho^2})\over\sqrt{\varrho}}\right] 
+ e^{i{\pi\over
4}}e^{-{3\pi\lambda\over 4}}\mid \varrho\mid^{-i{\lambda\over
2}}\,\left[ {\exp({+i\varrho^2})\over\sqrt{\varrho}}\right] \,,
\end{eqnarray}
\end{mathletters}
therefore, if we identify

\begin{equation}
f_1(x) = {\exp({-i\varrho^2})\over\sqrt{\varrho}}\qquad 
{\rm and}\qquad
       f_2(x) = {\exp({+i\varrho^2})\over\sqrt{\varrho}}\,,
\label{eqf12}
\end{equation}
we can conclude that the elements of {\bf A}$(+\infty)$-matrix will be

\begin{mathletters}
\label{eqampi}
\begin{eqnarray}
A_{11}(+\infty) &=&  e^{i{\pi\over 4}}e^{-{\pi\lambda\over 4}}
                          \mid \varrho\mid^{i{\lambda\over2}}\\
       A_{12}(+\infty) &=&  e^{-i{\pi\over 4}}e^{-{3\pi\lambda\over 4}}
                          \mid \varrho\mid^{i{\lambda\over2}}\\
       A_{21}(+\infty) &=&  e^{-i{\pi\over 4}}e^{-{\pi\lambda\over 4}}
                          \mid \varrho\mid^{-i{\lambda\over2}}\\
       A_{22}(+\infty) &=&  e^{i{\pi\over 4}}e^{-{3\pi\lambda\over 4}}
                          \mid \varrho\mid^{-i{\lambda\over2}}\,.
\end{eqnarray}
\end{mathletters}
In the case of $x\rightarrow -\infty$ if we consider that
\begin{equation}
\varrho^{\pm i{\lambda\over 2}} = e^{\mp (n+{1\over 2})\pi\lambda}
\mid \varrho\mid^{\pm i{\lambda\over 2}}\,,\qquad n = 0,1,2,\ldots\,,
\label{eqm1}
\end{equation}
in the Eqs.~(\ref{eqpas2}) then we can write two asymptotic solutions
of the Schr\"odinger equation when $x\rightarrow -\infty$ in the form

\begin{eqnarray}
\Psi_1(x\rightarrow -\infty) &=& \left[( e^{i{3\pi\over
4}} + e^{i{\pi\over 4}})e^{{\pi\lambda\over 4}} + e^{-i{\pi\over
4}}e^{-{\pi\lambda\over 4}}\right] \mid \varrho\mid^{i{\lambda\over
2}}\,\left[{\exp({-i\varrho^2})\over\sqrt{\varrho}}\right] 
+ \nonumber\\
&+& \left[( e^{i{3\pi\over
4}} + e^{i{\pi\over 4}})e^{{\pi\lambda\over 4}} - e^{i{\pi\over
4}}e^{-{\pi\lambda\over 4}}\right] \mid \varrho\mid^{-i{\lambda\over
2}}\,\left[ {\exp({+i\varrho^2})\over\sqrt{\varrho}}\right]
\label{eqpmai1}
\end{eqnarray}
and
\begin{eqnarray}
\Psi_2(x\rightarrow -\infty) &=& \left[( e^{i{3\pi\over
4}} + e^{i{\pi\over 4}})e^{-{\pi\lambda\over 4}} + e^{i{\pi\over
4}}e^{-{3\pi\lambda\over 4}}\right] \mid \varrho\mid^{i{\lambda\over
2}}\,\left[ {\exp ({-i\varrho^2})\over\sqrt{\varrho}}\right] 
+ \nonumber\\
&+& \left[( e^{i{3\pi\over
4}} + e^{i{\pi\over 4}})e^{-{\pi\lambda\over 4}} - e^{-i{\pi\over
4}}e^{-{3\pi\lambda\over 4}}\right] \mid \varrho\mid^{-i{\lambda\over
2}}\,\left[ {\exp ({+i\varrho^2})\over\sqrt{\varrho}}\right]\,,
\label{eqpmai2}
\end{eqnarray}
therefore we can identify the elements of {\bf A}$(-\infty)$-matrix
as

\begin{mathletters}
\label{eqammpi}
\begin{eqnarray}
A_{11}(-\infty) &=&  \left[( e^{i{3\pi\over
4}} + e^{i{\pi\over 4}})e^{{\pi\lambda\over 4}} + e^{-i{\pi\over
4}}e^{-{\pi\lambda\over 4}}\right] \mid \varrho\mid^{i{\lambda\over
2}}\\
       A_{12}(-\infty) &=& \left[( e^{i{3\pi\over
4}} + e^{i{\pi\over 4}})e^{-{\pi\lambda\over 4}} + e^{i{\pi\over
4}}e^{-{3\pi\lambda\over 4}}\right] \mid \varrho\mid^{i{\lambda\over
2}}\\
       A_{21}(-\infty) &=& \left[( e^{i{3\pi\over
4}} + e^{i{\pi\over 4}})e^{{\pi\lambda\over 4}} - e^{i{\pi\over
4}}e^{-{\pi\lambda\over 4}}\right] \mid \varrho\mid^{-i{\lambda\over
2}}\\
       A_{22}(-\infty) &=& \left[( e^{i{3\pi\over
4}} + e^{i{\pi\over 4}})e^{-{\pi\lambda\over 4}} - e^{-i{\pi\over
4}}e^{-{3\pi\lambda\over 4}}\right] \mid \varrho\mid^{-i{\lambda\over
2}}\,.
\end{eqnarray}
\end{mathletters}
In the choice of the two asymptotic wave functions for
$x\rightarrow\pm\infty$ we considered the set of properties given by
the Eqs.~(\ref{eqpfpr}) and (\ref{eqfa16}) that the {\bf F}-matrix
need to satisfies. Using the results for the {\bf A}$(-\infty)$
and {\bf A}$(+\infty)$ in the Eqs.~(\ref{eqfa9}) we can show that the
evolution matrix can be written as

\begin{equation}
{\bf F}(-\infty,+\infty ) = \left[ \matrix{ie^{\pi\lambda\over 2}&
(1+ie^{\pi\lambda\over 2})\mid \varrho\mid^{i\lambda}\cr 
(-1+ie^{\pi\lambda\over 2})\mid 
\varrho\mid^{-i\lambda}&ie^{\pi\lambda\over 2}
\cr}\right]
\label{eqmfpf}
\end{equation}
and the exact transmission and reflection coefficients are given by

\begin{equation}
T = {1\over \mid F_{12}(-\infty,+\infty)\mid^2} = {1\over
1+e^{\pi\lambda}}\,,
\label{eqtcp}
\end{equation}
and

\begin{equation}
R = {\mid F_{22}(-\infty,+\infty)\mid^2\over \mid 
F_{12}(-\infty,+\infty)\mid^2} = {e^{\pi\lambda}\over
1+e^{\pi\lambda}}\,.
\label{eqrcp}
\end{equation}

\subsection{Morse Barrier}

For a Morse potential barrier\cite{refbellm}

\begin{equation}
V_1(x) = V_0\left( 2e^{x/b} - e^{2x/b}\right)\,,
\label{eqmb1}
\end{equation}
the corresponding superpotential, obtained by the Eq.~(\ref{eqv1}),
is given by

\begin{equation}
W(x,a_1) = a_1 + \alpha\,e^{x/b} 
\label{eqwm1}
\end{equation}
where 

\begin{equation}
\cases{a_1 = \sqrt{\varepsilon}\left(1\mp is\right)\cr
             \alpha = \pm i\sqrt{V_0}\cr}\,,
\label{eqcwm1}
\end{equation}
with \ $\varepsilon = \hbar^2/(8mb^2)$\ and\ $s =
\sqrt{V_0/\varepsilon}$.  The shape invariance condition
(\ref{eqvv12}) imply that

\begin{equation}
R(a_n) = a_n^2-a_{n+1}^2\,,
\label{eqranm}
\end{equation}
where $a_{n+1} = a_n+2\sqrt\varepsilon$. Using the Eq.~(\ref{eqllda})
we can conclude that

\begin{equation}
\sum_{k=1}^\mu R(a_k) = a_1^2 - a_{\mu+1}^2 = \Lambda = E + a_1^2
\label{eqrlm}
\end{equation}
or

\begin{equation}
a_{\mu+1} = \pm i\sqrt{E}\,.
\label{eqamum}
\end{equation}
If we remember that
\begin{equation}
\mu = {a_{\mu+1}-a_1\over 2\sqrt{\varepsilon}}\,,
\label{eqmum}
\end{equation}
we can use the Eqs.~(\ref{eqcwm1}) and (\ref{eqamum}) to show that

\begin{equation}
\mu = -{1\over 2}\pm i{s\over 2}\pm i{r\over 2}\,,
\label{eqmum1}
\end{equation}
where $r = \sqrt{E/\varepsilon}$.

Considering the asymmetry of the Morse potential barrier the
wave function will have a different behavior in
$+\infty$ and $-\infty$. Therefore the asymptotic form of the
components of the wave function for $x\rightarrow +\infty$ can be
obtained using the last results in the Eqs.~(\ref{eqpssl})

\begin{mathletters}
\label{eqmas}
\begin{eqnarray}
\Psi_-(x\rightarrow +\infty) &=& \beta 
\,e^{\mp i{\pi\over 4}}\,e^{-{\pi\over
4}\left(s\pm r\right)}\,\left[ {\exp{\left(\mp i{s\over
2}\exp(x/b)\right)}\over\sqrt{\exp(x/b)}}\right]\\
\Psi_+(x\rightarrow +\infty) &=& 
\gamma\,e^{\mp i{\pi\over 4}}\,e^{{\pi\over
4}\left(s\pm r\right)}\,\left[ {\exp{\left(\pm i{s\over 2}\exp(x/b)
\right)}\over\sqrt{\exp(x/b)}}\right]\,.
\end{eqnarray}
\end{mathletters}
Using these results it is possible to write two asymptotic solutions
of the Schr\"odinger equation when $x\rightarrow +\infty$ in the form

\begin{mathletters}
\label{eqpmai}
\begin{eqnarray}
\Psi_1(x\rightarrow +\infty) = e^{i{\pi\over
4}}e^{-{\pi\over 4}\left(s+r\right)}\, &&\!\!\!\!\!\!\left[ 
{\exp{\left(-i{s\over 2} 
\exp({x/ b})\right)}\over\sqrt{\exp({x/ b})}}\right]\\ \nonumber
&+&
e^{-i{\pi\over 4}}e^{{\pi\over 4}\left(s-r\right)}\,\left[
{\exp{\left(+i{s\over 2}\exp({x/ b})\right)}\over\sqrt{\exp ({x/
b})}}\right] \\
\Psi_2(x\rightarrow +\infty) = e^{-i{\pi\over
4}}e^{-{\pi\over 4}\left(s-r\right)}\,&&\!\!\!\!\!\! 
\left[ {\exp{\left(-i{s\over 2} 
\exp({x/ b})\right)}\over\sqrt{\exp({x/ b})}}\right]\\ \nonumber
&+&
e^{i{\pi\over 4}}e^{{\pi\over 4}\left(s+r\right)}\,\left[
{\exp{\left(+i{s\over 2}\exp({x/ b})\right)}\over\sqrt{\exp ({x/
b})}}\right]\,,
\end{eqnarray}
\end{mathletters}
therefore, if we identify

\begin{equation}
f_1(x\rightarrow +\infty) = {\exp{\left(-i{s\over 2}\exp({x/
b})\right)}\over\sqrt{\exp({x/ b}})}\qquad {\rm and}\qquad
f_2(x\rightarrow +\infty) = {\exp{\left(+i{s\over 2}\exp ({x/
b})\right)}
\over\sqrt{\exp ({x/ b})}}\,,
\label{eqf12m}
\end{equation}
we can conclude that the elements of {\bf A}$(+\infty)$-matrix will be

\begin{mathletters}
\label{eqammi}
\begin{eqnarray}
A_{11}(+\infty) &=&  e^{i{\pi\over 4}}e^{-{\pi\over 
4}\left(s+r\right)}\\
A_{12}(+\infty) &=&  e^{-i{\pi\over 4}}e^{-{\pi\over 
4}\left(s-r\right)}\\
A_{21}(+\infty) &=&  e^{-i{\pi\over 4}}e^{{\pi\over 
4}\left(s-r\right)}\\
A_{22}(+\infty) &=&  e^{i{\pi\over 4}}e^{{\pi\over 
4}\left(s+r\right)}\,.
\end{eqnarray}
\end{mathletters}
In the case of $x\rightarrow -\infty$ we can substitute for $s$ and
$r$ and using the Eq.~(\ref{eqpssf}) find

\begin{mathletters}
\label{eqmes}
\begin{eqnarray}
\Psi_-(x\rightarrow -\infty) &=& \beta \,{\sqrt{r}\,\Gamma(\pm
ir)\over \Gamma\left({1\over 2}\pm i{s\over 2}\pm i{r\over 2}\right)}\,
e^{\mp ikx}\\
\Psi_+(x\rightarrow -\infty) &=& \gamma\,{\Gamma\left({1\over
2}\pm i{s\over 2}\pm i{r\over 2}\right)\over \sqrt{r}\,\Gamma(\pm
ir)}\,e^{\pm ikx}\,,
\end{eqnarray}
\end{mathletters}
where \ $k = \sqrt{2mE}/\hbar$.  Using these results it is possible to
write two asymptotic solutions of the Schr\"odinger equation when
$x\rightarrow -\infty$ in the form

\begin{eqnarray}
\Psi_1(x\rightarrow -\infty) &=& \left[{e^{i{3\pi\over
4}}\,e^{-{\pi r\over 2}}\over \Gamma\left({1\over 2}+i{s\over
2}+i{r\over 2}\right)}+{e^{i{\pi\over
4}}\,e^{{\pi s\over 2}}\over \Gamma\left({1\over 2}+i{s\over
2}-i{r\over 2}\right)}\right]\sqrt{r}\,\Gamma\left(ir\right)\,e^{ikx} 
\nonumber\\
&+& \left[{e^{i{\pi\over
4}}\,e^{{\pi \over 2}(s-r)}\over \Gamma\left({1\over 2}-i{s\over
2}-i{r\over 2}\right)}+{e^{i{3\pi\over
4}}\over \Gamma\left({1\over 2}-i{s\over
2}+i{r\over 2}\right)}\right]\sqrt{r}\,\Gamma\left(ir\right)\,e^{-ikx}
\,,
\label{eqpmm1}
\end{eqnarray}
and

\begin{eqnarray}
\Psi_2(x\rightarrow -\infty) &=& \left[{e^{i{\pi\over
4}}\over \Gamma\left({1\over 2}+i{s\over
2}+i{r\over 2}\right)}+{e^{i{3\pi\over
4}}\,e^{{\pi\over 2}(s+r)}\over \Gamma\left({1\over 2}+i{s\over
2}-i{r\over 2}\right)}\right]\sqrt{r}\,\Gamma\left(-ir\right)\,e^{ikx}
 \nonumber\\
&+& \left[{e^{i{\pi\over
4}}\,e^{{\pi r\over 2}}\over \Gamma\left({1\over 2}-i{s\over
2}-i{r\over 2}\right)}+{e^{i{3\pi\over
4}}\,e^{\pi s\over 2}\over \Gamma\left({1\over 2}-i{s\over
2}-i{r\over
2}\right)}\right]\sqrt{r}\,\Gamma\left(-ir\right)\,e^{-ikx} \,,
\label{eqpmm2}
\end{eqnarray}
therefore, if we identify

\begin{equation}
f_1(x\rightarrow -\infty) = {\exp{(+ikx)}\over\sqrt{k}}\qquad {\rm 
and}\qquad
      f_2(x\rightarrow -\infty) = {\exp{(-ikx)}\over\sqrt{k}}\,,
\label{eqf12mme}
\end{equation}
we can conclude that the elements of {\bf A}$(-\infty)$-matrix will be

\begin{mathletters}
\label{eqamm12}
\begin{eqnarray}
A_{11}(-\infty) &=& \left[{e^{i{3\pi\over
4}}\,e^{-{\pi r\over 2}}\over \Gamma\left({1\over 2}+i{s\over
2}+i{r\over 2}\right)}+{e^{i{\pi\over
4}}\,e^{{\pi s\over 2}}\over \Gamma\left({1\over 2}+i{s\over
2}-i{r\over 2}\right)}\right]\sqrt{r}\,\Gamma\left(ir\right)\\
A_{12}(-\infty) &=&  \left[{e^{i{\pi\over
4}}\over \Gamma\left({1\over 2}+i{s\over
2}+i{r\over 2}\right)}+{e^{i{3\pi\over
4}}\,e^{{\pi\over 2}(s+r)}\over \Gamma\left({1\over 2}+i{s\over
2}-i{r\over 2}\right)}\right]\sqrt{r}\,\Gamma\left(-ir\right)\\
A_{21}(-\infty) &=& \left[{e^{i{\pi\over
4}}\,e^{{\pi \over 2}(s-r)}\over \Gamma\left({1\over 2}-i{s\over
2}-i{r\over 2}\right)}+{e^{i{3\pi\over
4}}\over \Gamma\left({1\over 2}-i{s\over
2}+i{r\over 2}\right)}\right]\sqrt{r}\,\Gamma\left(ir\right)\\
A_{22}(-\infty) &=& \left[{e^{i{\pi\over
4}}\,e^{{\pi r\over 2}}\over \Gamma\left({1\over 2}-i{s\over
2}-i{r\over 2}\right)}+{e^{i{3\pi\over
4}}\,e^{\pi s\over 2}\over \Gamma\left({1\over 2}-i{s\over
2}-i{r\over 2}\right)}\right]\sqrt{r}\,\Gamma\left(-ir\right)\,.
\end{eqnarray}
\end{mathletters}
Again, in the choice of the two asymptotic wave functions for
$x\rightarrow\pm\infty$ we have considered the properties of the {\bf
F}-matrix. Using the results for the {\bf A}$(-\infty)$ and {\bf
A}$(+\infty)$ in the Eqs.~(\ref{eqfa9}) we can show that the evolution
matrix can be written as

\begin{equation}
{\bf F}(-\infty,+\infty ) = \left[ \matrix{ig&ih\cr 
ih^* &ig^*\cr}\right]
\label{eqmfpfm}
\end{equation}
where

\begin{equation}
g = {e^{{\pi\over 4}\left(s-r\right)}\sqrt{r}\,\Gamma(ir)\over
\Gamma\left({1\over 2}+i{s\over 2}+i{r\over 2}\right)}
\label{eqmbag}
\end{equation}
and 

\begin{equation}
h = {e^{{\pi\over 4}\left(s+r\right)}\sqrt{r}\,\Gamma(ir)\over
\Gamma\left({1\over 2}+i{s\over 2}-i{r\over 2}\right)}\,.
\label{eqmbah}
\end{equation}
In this case the exact transmission and reflection
coefficients are given by

\begin{equation}
T = {1\over \mid F_{12}(-\infty,+\infty)\mid^2} = {e^{-{\pi\over
2}\left(s+r\right)}\sinh{(\pi r)}\over 
\cosh{\left[{\pi\over 2}\left(s-r\right)\right]}}\,,
\label{eqtcm}
\end{equation}
and

\begin{equation}
R = {\mid F_{22}(-\infty,+\infty)\mid^2\over \mid
F_{12}(-\infty,+\infty)\mid^2} = {e^{-\pi r}\cosh{\left[{\pi\over
2}\left(s+r\right)\right]}\over\cosh{\left[{\pi\over
2}\left(s-r\right) \right]}}\,.
\label{eqrcm}
\end{equation}

\subsection{Eckart Barrier}

For an Eckart potential barrier\cite{refeckart}

\begin{equation}
V_1(x) = V_0\, {\rm sech}^2{\left(x/2b\right)}\,,
\label{eqeb1}
\end{equation}
the correspondent superpotential, obtained by the Eq.~(\ref{eqv1}),
is given by

\begin{equation}
W(x,a_1) = a_1\tanh{\left(x/2b\right)}
\label{eqwe1}
\end{equation}
where 

\begin{equation}
a_1 = \sqrt{\varepsilon}\left(-1\pm is\right)\,,
\label{eqcwe1}
\end{equation}
with \  $\varepsilon = \hbar^2/(32mb^2)$\  and\  
$s = \sqrt{V_0/\varepsilon-1}$.
The shape invariance condition (\ref{eqvv12}) imply that

\begin{equation}
R(a_n) = a_n^2-a_{n+1}^2\,,
\label{eqrane}
\end{equation}
where $a_{n+1} = a_n-2\sqrt\varepsilon$. Using the Eq.~(\ref{eqllda})
we can conclude that

\begin{equation}
\sum_{k=1}^{2\nu} R(a_k) = a_1^2 - a_{2\nu+1}^2 = \Lambda = E + a_1^2
\label{eqrle}
\end{equation}
or

\begin{equation}
a_{2\nu+1} = \pm i\sqrt{E}\,.
\label{eqmue}
\end{equation}
If we remember that
\begin{equation}
2\nu = {a_{2\nu+1}-a_1\over -2\sqrt{\varepsilon}}\,,
\label{eqmuec}
\end{equation}
we can use the Eqs.~(\ref{eqcwe1}) and (\ref{eqmue}) to show that

\begin{equation}
2\nu = -{1\over 2}\pm i{s\over 2}\mp i{r\over 2}\,,
\label{eqmue1}
\end{equation}
where $r = \sqrt{E/\varepsilon}$.

The asymptotic form of the components of the wave function can be
obtained using these results in the Eqs.~(\ref{eqpsb})

\begin{mathletters}
\label{eqeas}
\begin{eqnarray}
\Psi_-(x) &=& \beta \,{\Gamma\left({3\over 4}\pm i{s\over 4}\pm
i{r\over 4}\right)\over \Gamma\left(1\pm i{r\over
2}\right)\,\Gamma\left({1\over 4}\pm i{s\over 4}\mp i{r\over
4}\right)}\, e^{\mp ikx}\\
\Psi_+(x) &=& \gamma\,{\Gamma\left(\pm i{r\over
2}\right)\,\Gamma\left({1\over 4}\pm i{s\over 4}\mp i{r\over
4}\right)\over \Gamma\left({3\over 4}\pm i{s\over 4}\pm i{r\over
4}\right)}\, e^{\pm ikx} \,,
\end{eqnarray}
\end{mathletters}
where \ $k = \sqrt{2mE}/\hbar$. Using these results and the relation 

\begin{equation}
\Gamma\left({1\over 4}\pm iy\right)\,\Gamma\left({3\over 4}\mp
iy\right) = {\sqrt{2}\pi\over \cosh{(\pi y)}\pm\sinh{(\pi y)}}\,,
\label{eqgaux}
\end{equation}
it is possible to write two asymptotic solutions of the Schr\"odinger
equation when $x\rightarrow +\infty$ in the form 

\begin{mathletters}
\label{eqpeai+}
\begin{eqnarray}
\Psi_1(x\rightarrow +\infty) &=& C_1e^{-ikx}+C_1^*e^{ikx}\\
\Psi_2(x\rightarrow +\infty) &=& C_2^*e^{-ikx}-C_2e^{ikx}\,,
\end{eqnarray}
\end{mathletters}
where

\begin{eqnarray}
C_1 &=& \sqrt{2}\,\pi\left\{ \cosh{\left[{\pi\over 4}\left(r\pm
s\right)\right] } + i\sinh{\left[{\pi\over 4}\left(r\pm
s\right)\right] }\right\}^{-1}\, \nonumber\\ &\times& 
\left\{ \Gamma\left(1+i{r\over
2}\right)\,\Gamma\left({1\over 4}\pm i{s\over 4}-i{r\over
4}\right)\,\Gamma\left({1\over 4}\mp i{s\over 4}-i{r\over 4}
\right)\right\}^{-1}
\label{eqc1}
\end{eqnarray}
and

\begin{eqnarray}
C_2 &=& \sqrt{2}\,\pi\,\Gamma\left(i{r\over 2}\right)\,\left\{ 
\cosh{\left[{\pi\over 4}\left(r\pm
s\right)\right] } - i\sinh{\left[{\pi\over 4}\left(r\pm
s\right)\right] }\right\}^{-1}\, \nonumber\\ &\times& \left\{
\Gamma\left({3\over 4}\mp i{s\over 4}+i{r\over
4}\right)\,\Gamma\left({3\over 4}\pm i{s\over 4}+i{r\over 4}\right)
\right\}^{-1}\,.
\label{eqc2}
\end{eqnarray}
Therefore, if we identify in the Eq.~(\ref{eqpeai+})

\begin{equation}
f_1(x\rightarrow +\infty) = {e^{-ikx}\over \sqrt{k}}\qquad {\rm and}\qquad
f_2(x\rightarrow +\infty) = {e^{+ikx}\over \sqrt{k}}\,,
\label{eqfe12+}
\end{equation}
we can conclude that the elements of {\bf A}$(+\infty)$-matrix will be

\begin{mathletters}
\label{eqamei+}
\begin{eqnarray}
A_{11}(+\infty) &=&  C_1\\
A_{12}(+\infty) &=&  C_2^*\\
A_{21}(+\infty) &=&  C_1^*\\
A_{22}(+\infty) &=&  -C_2\,.
\end{eqnarray}
\end{mathletters}
Also we can write two asymptotic solutions of
the Schr\"odinger when $x\rightarrow -\infty$ in the form

\begin{mathletters}
\label{eqpeai-}
\begin{eqnarray}
\Psi_1(x\rightarrow -\infty) &=& C_1e^{ikx}-C_1^*e^{-ikx}\\
\Psi_2(x\rightarrow -\infty) &=& -C_2^*e^{ikx}-C_2e^{-ikx}\,,
\end{eqnarray}
\end{mathletters}
and identifying

\begin{equation}
f_1(x\rightarrow -\infty) = {e^{+ikx}\over 
\sqrt{k}}\qquad {\rm and}\qquad
f_2(x\rightarrow -\infty) = {e^{-ikx}\over \sqrt{k}}\,,
\label{eqfe12-}
\end{equation}
we can conclude that the elements of {\bf A}$(-\infty)$-matrix will be

\begin{mathletters}
\label{eqamei-}
\begin{eqnarray}
A_{11}(-\infty) &=&  C_1\\
A_{12}(-\infty) &=&  -C_2^*\\
A_{21}(-\infty) &=&  -C_1^*\\
A_{22}(-\infty) &=&  -C_2\,.
\end{eqnarray}
\end{mathletters}
Using the results for the {\bf A}$(-\infty)$ and {\bf A}$(+\infty)$ in
the Eq.~(\ref{eqfa9}) we can show that the evolution matrix can be
written as

\begin{equation}
{\bf F}(-\infty,+\infty ) = \left[ \matrix{g&h\cr -h^* &-g^*\cr}\right]
\label{eqmfpfe}
\end{equation}
where

\begin{equation}
g = {C_1C_2-C_1^*C_2^*\over C_1C_2+C_1^*C_2^*}
\label{eqebag}
\end{equation}
and 

\begin{equation}
h = {2C_1C_2^*\over C_1C_2+C_1^*C_2^*}\,.
\label{eqebah}
\end{equation}
On using the Eqs.~(\ref{eqc1}), (\ref{eqc2}) after a considerable
amount of algebra we can show that the exact transmission and
reflection coefficients are given by

\begin{equation}
T = {1\over \mid F_{12}(-\infty,+\infty)\mid^2} 
= {1\over 4}\,\Biggl\vert
{C_2\over C_2^*} + {C_1^*\over C_1}\Biggr\vert ^2 = 
{\sinh^2{\left({\pi
r\over 2}\right)}\over \sinh^2{\left({\pi r\over 2}\right)} +
\cosh^2{\left({\pi s\over 2}\right)}}\,,
\label{eqtce}
\end{equation}
and

\begin{equation}
R = {\mid F_{22}(-\infty,+\infty)\mid^2\over \mid
F_{12}(-\infty,+\infty)\mid^2} =  {1\over 4}\,\Biggl\vert
{C_2\over C_2^*} - {C_1^*\over C_1}\Biggr\vert ^2 
= {\cosh^2{\left({\pi
s\over 2}\right)}\over \sinh^2{\left({\pi r\over 2}\right)} +
\cosh^2{\left({\pi s\over 2}\right)}}\,.
\label{eqrce}
\end{equation}

\section{Concluding remarks}

In conclusion we note that the present technique is a powerful and
an elegant prescription to obtain exact reflection and transmission
coefficients. This method may also be used for all supersymmetric
shape invariant potential barriers that satisfy the analytic
continuation condition (\ref{eqwz}).

One possible application of shape-invariance formalism is to
multidimensional quantum tunneling. In nuclear physics applications
multidimensional quantum tunneling can be visualized as the tunneling
of a quantum mechanical system (such as a nucleus with internal
excitation) instead of a structureless particle through a
one-dimensional barrier. The nucleus typically taken to enter the
barrier in its ground state and may emerge either in the ground state
or in an excited state at the other side of the barrier. The
interaction between the penetrating quantum system and the barrier
also needs to be specified based on the physical conditions of the
problem. It has been known for some time that, if the excitation
energies are neglected, the penetration probability of an
$N$-dimensional system can be reduced to a sum of probabilities of $N$
one-dimensional suitably defined barriers \cite{zeropoint}. The
eigenchannel formulation remains valid even for finite excitation
energies as long as the energy-dependence of the weight factors is
taken into account \cite{hagino}. Our formulation would be applicable
in such cases if the eigenpotentials are shape-invariant. A simpler
limit would assume factorization of the interaction between the
barrier and the quantum system into a product of two quantities which
are functions of the barrier and internal degrees of freedom
respectively. Such a factorization approach was already applied to a
coupled system of equations for bound states \cite{daschak}.

Our formulation also casts the tunneling problem in an algebraic basis
\cite{ref3,asim}. If the internal system can be described by an
algebraic model such as the interacting boson model \cite{ibmfus} then
it may be possible to cast the entire problem into an algebraic
framework. A group-theoretical formulation can be a starting
point of systematic approximations such as those given in Ref. 
\cite{nag}. A detailed study of such aspects is deferred to later
work.

\section*{ACKNOWLEDGMENTS}

This work was supported in part by the U.S. National Science
Foundation Grant No.\ PHY-9605140 at the University of Wisconsin, and
in part by the University of Wisconsin Research Committee with funds
granted by the Wisconsin Alumni Research Foundation.  M.A.C.R.\ 
acknowledges the support of Funda\c c\~ao de Amparo \`a Pesquisa do
Estado de S\~ao Paulo (Contract No.\ 98/13722-2). A.N.F.A.
acknowledges the support of Funda\c c\~ao Coordena\c c\~ao de
Aperfei\c coamento de Pessoal de N\'{\i}vel Superior (Contract No.
BEX0610/96-8). M.A.C.R. thanks to the Nuclear Theory Group at
University of Wisconsin for the very kind hospitality.  We also thank
the Institute for Nuclear Theory at the University of Washington for
its hospitality and Department of Energy for partial support during
the early stages of this work.

\end{document}